\documentclass[12pt]{article}
\usepackage{a4}
\usepackage{amssymb}
\usepackage{cite}
\newcommand{\be}{\begin{equation}}
\newcommand{\ee}{\end{equation}}
\newcommand{\bea}{\begin{eqnarray}}
\newcommand{\eea}{\end{eqnarray}}

\begin{document}
\thispagestyle{empty}
\def\thefootnote{\fnsymbol{footnote}}
%\begin{flushright}
%{\sf ZMP-HH/09-17}\\%[1mm]
%{\sf Hamburger$\;$Beitr\"age$\;$zur$\;$Mathematik$\;$Nr.$\;$ 345}
%     \vskip 2em
%\end{flushright}
%\vskip 2.0em
\vspace*{1cm}
\begin{center}\Large
On canonical quantization of  the gauged WZW model with permutation branes
\end{center}\vskip 1.5em
\begin{center}
Gor Sarkissian
\footnote{\scriptsize
~Email address: \\
$~$\hspace*{2.4em} gor.sarkissian@ysu.am
}
\end{center}
\begin{center}
Department of Theoretical Physics, \ Yerevan State University,\\
Alex Manooghian 1, \  Yerevan 0025,\\
Armenia
\end{center}
\vskip 1.5em
\begin{center} February 2011 \end{center}
\vskip 2em
\begin{abstract} \noindent
In this paper we perform canonical quantization of the product of the gauged WZW models on a strip
with boundary conditions specified by permutation branes. We show that the phase space
of the $N$-fold product of the gauged WZW model $G/H$  on a strip with boundary conditions given by permutation branes is symplectomorphic to the phase space of the double Chern-Simons theory on a sphere  with $N$ holes times 
the time-line with $G$ and $H$ gauge fields both coupled to two 
Wilson lines. For the special case of the topological coset $G/G$  we arrive at the conclusion
that  the phase space
of the $N$-fold product of the topological coset $G/G$  on a strip with boundary conditions given by permutation branes is symplectomorphic to the phase space of  Chern-Simons theory on a Riemann surface of the genus $N-1$
times the time-line with four Wilson lines.
 \end{abstract}
\vspace{0.5cm}

\setcounter{footnote}{0}
\def\thefootnote{\arabic{footnote}}
\newpage
%%%%%%%%%%%%%%%%%%%%%%%%%%%%%%%%%%%%%%%%%%%%%%%%%%%%%%%%%%%%%%%%%%%%%%%%

\section{Introduction}
In this paper we continue to investigate  the problem of the canonical quantization of the WZW and gauged WZW
models with defects and branes, started in the papers \cite{Sarkissian:2009hy,Sarkissian:2010bw}.
In the paper \cite{Sarkissian:2009hy} we addressed the problem of the canonical quantization of the WZW
model with defects and permutation branes.  In the paper   \cite{Sarkissian:2010bw} the canonical
quantization of the gauged WZW model with defects has been performed. In the present
paper we turn to the canonical quantization of the gauged WZW model $G/H$ with permutation branes.
It was shown in the paper  \cite{Sarkissian:2009hy} that the phase space of the $N$-fold product of  WZW models
on a strip with boundary conditions given by permutation branes is symplectomorphic to the phase space of
Chern-Simons theory on a sphere with $N$ holes times the time-line with two Wilson lines. Using the ansatz for the permutation branes on product of cosets suggested in \cite{Sarkissian:2006xp}, here we show that the phase space
of the $N$-fold product of the gauged WZW models on a strip with boundary conditions given by permutation branes is symplectomorphic to the phase space of
the double Chern-Simons theory on a sphere with $N$ holes times the time-line with $G$ and $H$ gauge fields both coupled to two 
Wilson lines.  For the case of the topological coset $G/G$ we get
that the phase space
of the $N$-fold product of the topological coset on a strip with boundary conditions given by permutation branes is symplectomorphic to the phase space of
the  Chern-Simons theory on a Riemann surface of the genus $N-1$ times the time-line with four Wilson lines.

\section{Bulk WZW model}
In this section we review the canonical quantization of the
WZW model with compact, simple, connected and simply connected group $G$  on the cylinder $\Sigma=R\times S^1=(t,x\; {\rm mod}\; 2\pi)$ 
\cite{Chu:1991pn,Falceto:1992bf,Gawedzki:1990jc}. 
The world-sheet action of the bulk WZW model is \cite{Witten:1983ar}
\bea
S^{\rm WZW}(g)&=&{k \over 4\pi}\int_{\Sigma}{\rm Tr}(g^{-1}\partial_{+} g)(g^{-1}\partial_{-}g)dx^+dx^-
 +{k \over 4\pi}\int_B {1\over 3}{\rm tr}(g^{-1}d g)^3\\ \nonumber
&\equiv& { k\over{4 \pi}} \left[ \int_{\Sigma}dx^+dx^- L^{\rm kin} 
+ \int_B \omega^{\rm WZ}\right] \, ,
 \eea
where $x^{\pm}=x\pm t$.
%\be
%\omega^{WZ}(g)={1\over 3}{\rm tr}(g^{-1}d g)^3
%\ee
The phase space of solutions ${\cal P}$ can be described by the Cauchy data 
\footnote{ Surely we can choose any time slice, but for simplicity we always below take the slice $t=0$.}
at $t=0$:
\be
g(x)=g(0,x)\;\;\; {\rm and}\;\;\; \xi_0(x)=g^{-1}\partial_tg(0,x)\, .
\ee

The corresponding symplectic form is \cite{Gawedzki:1990jc}:
\be
\Omega^{\rm bulk}={k\over 4\pi}\int_0^{2\pi}\Pi^{G}(g) dx\, ,
\ee
where 
\be\label{pigden}
\Pi^{G}(g)={\rm tr}\left(-\delta\xi_0g^{-1}\delta g+
(\xi_0+g^{-1}\partial_x g)(g^{-1}\delta g)^2\right)\, .
\ee
The $\delta$ denotes here the exterior derivative on the phase space ${\cal P}$.
It is easy to check that the symplectic form density $\Pi(g)$ has the following exterior derivative
\be\label{dpi}
\delta \Pi^{G}(g)=\partial_x\omega^{WZ}(g)\, ,
\ee
what implies closedness of the $\Omega$:
\be
\delta\Omega^{\rm bulk}=0\, .
\ee

The classical equations of motion are
\be\label{eqmot}
\partial_{-}J_L=0\;\;\;{\rm and} \;\;\;\partial_{+}J_R=0\, ,
\ee
where 
\be
J_L=-ik \partial_{+}gg^{-1}\;\;\;{\rm and}\;\;\; J_R=ikg^{-1}\partial_{-}g\, .
\ee
The general solution of (\ref{eqmot}) satisfying the boundary conditions:
\be
g(t,x+2\pi)=g(t,x)
\ee
is
\be\label{bulkdec}
g(t,x)=g_L(x^+)g^{-1}_R(x^-)
\ee
with $g_{L,R}$ satisfying the monodromy conditions:
\be\label{monl}
g_{L}(x^++2\pi)=g_{L}(x^+)\gamma\, ,
\ee
\be\label{monr}
g_{R}(x^-+2\pi)=g_{R}(x^-)\gamma
\ee
with the same matrix $\gamma$.
Expressing the symplectic form density $\Pi^{G}(g)$ in the terms of $g_{L,R}$ we obtain:
\be\label{denlr}
\Pi^G=
{\rm tr}\left[g_L^{-1}\delta g_L\partial_x(g_L^{-1}\delta g_L)-
g_R^{-1}\delta g_R\partial_x(g_R^{-1}\delta g_R)
+\partial_x (g_L^{-1}\delta g_Lg_R^{-1}\delta g_R)\right]\, .
\ee
Using (\ref{denlr}) and (\ref{monl}), (\ref{monr}) one derives for $\Omega^{\rm bulk}$:
\be\label{rldeco}
\Omega^{\rm bulk}=\Omega^{\rm chiral}(g_L,\gamma)-\Omega^{\rm chiral}(g_R, \gamma)\, ,
\ee
where
\be
\Omega^{\rm chiral}(g_L,\gamma)={k\over 4\pi}\int_0^{2\pi}{\rm tr}\left(g_L^{-1}\delta g_L\partial_x(g_L^{-1}\delta g_L)\right)dx+
{k\over 4\pi}{\rm tr}(g_L^{-1}\delta g_L(0)\delta\gamma\gamma^{-1})\, .
\ee
%and $\Omega_R$ is given by the same formula with $g_R\rightarrow g_L$.
The chiral field $g_L$ can be decomposed into the product of a closed loop in $G$,
a multivalued field in the Cartan subgroup and a constant element in $G$:
\be\label{decomp}
g_L(x^+)=h(x^+)e^{i\tau x^+/k}g_0^{-1}\, ,
\ee
where $h\in LG$, $\tau\in t$ ( the Cartan algebra) and $g_0\in G$. For the monodromy
of $g_L$ we find:
\be\label{mondecomp}
\gamma=g_0e^{2i\pi\tau/k }g_0^{-1}\, .
\ee
The parametrization (\ref{decomp}) induces the following decomposition of $\Omega^{\rm chiral}(g_L,\gamma)$:
\be\label{omdecom}
\Omega^{\rm chiral}(g_L,\gamma)=\Omega^{LG}(h,\tau)+{k\over 4\pi}\omega_{\tau}(\gamma)+{\rm tr}[(i\delta\tau)g_0^{-1}\delta g_0]\, ,
\ee
where $\Omega^{LG}(h,\tau)$ is :
\be\label{omlg}
\Omega^{LG}(h,\tau)={k\over 4\pi}\int_0^{2\pi}{\rm tr}[h^{-1}\delta h\partial_x(h^{-1}\delta h)+
{2i\over k}\tau(h^{-1}\delta h)^2
-{2i\over k}(\delta \tau)h^{-1}\delta h]dx
\ee
and $\omega_{\tau}(\gamma)$ is :
\be\label{omfik}
\omega_{\tau}(\gamma)={\rm tr}[g_0^{-1}\delta g_0e^{2i\pi\tau/k}g_0^{-1}\delta g_0e^{-2i\pi\tau/k}]\, .
\ee

Comparing (\ref{rldeco})  and (\ref{omlg}) to the formulae in appendix C  we see that the symplectic phase space of the 
WZW model on a cylinder coincides with that of Chern-Simons theory on the annulus times the time-line ${\cal A}\times R$.

\section{Gauged WZW model}
Here we review quantization of the gauged WZW model on the cylinder $\Sigma=R\times S^1=(t,x\; {\rm mod}\; 2\pi)$ 
as it is done in \cite{Gawedzki:2001ye}.

The action of the gauged WZW model is \cite{Bardakci:1987ee,Gawedzki:1988hq,Gawedzki:1988nj,Karabali:1988au}:
\be\label{gauact}
S^{G/H}(g,A)=S^{\rm WZW}+S^{\rm gauge}\, ,
\ee
where
\be
S^{\rm gauge}={k\over 2\pi }\int_{\Sigma}L^{\rm gauge}\, ,
\ee
\be
L^{\rm gauge}(g,A)=-{\rm tr}[-\partial_{+}gg^{-1}A_{-}+g^{-1}\partial_{-}gA_{+}+gA_{+}g^{-1}A_{-}-A_{+}A_{-}]\, .
\ee
With the help of  the Polyakov-Wiegmann identities:
\begin{eqnarray}
\label{pwk}
L^{\rm kin}(g h) &=&  L^{\rm kin}(g) + L^{\rm kin} (h)
 +  {\rm Tr} \big(g^{-1}\partial_z g \partial_{\bar z} h h^{-1}\big)+
 {\rm Tr} \big(g^{-1} \partial_{\bar z} g\partial_z  h h^{-1}\big)\, , \\[1ex]
\label{pwwz}
\omega^{\rm WZ}(g h) &=& \omega^{\rm WZ}(g) + \omega^{\rm WZ}(h)
 - {\rm d}\Big({\rm Tr} \big(g^{-1} {\rm d}g  {\rm d}h h^{-1}\big)\Big)\, ,
\end{eqnarray}
it is easy to check that the action (\ref{gauact}) is invariant under the gauge transformation:
\be
g\rightarrow hgh^{-1}\, ,\hspace{1cm} A\rightarrow hAh^{-1}-dhh^{-1}
\ee
for $h: \Sigma\rightarrow H$.

The equations of  motions are:

\be
\label{eom}
D_{+}(g^{-1}D_{-}g)=0\, ,\;\;\:  {\rm Tr}(g^{-1}D_{-}gT_H)={\rm Tr}(gD_{+}g^{-1}T_H)=0\, , \;\;\; F(A)=0\, ,
\ee
where $D_{\pm}g=\partial_{\pm}g+[A_{\pm},\, g]$  and $T_H$  is any element in the $H$ Lie algebra.

The flat gauge field  $A$ can be written as $h^{-1}dh$ for $h:\, R^2\rightarrow H$ and satisfying:
\be
h(t, x+2\pi)=\rho^{-1}h(t,x)
\ee
for some $\rho\in H$.

Define $\tilde{g}=hgh^{-1}$. Note that $\tilde{g}$ satisfies
\be\label{mongrho}
\tilde{g}(t,x+2\pi)=\rho^{-1}\tilde{g}(t,x)\rho\, .
\ee

In the terms of $\tilde{g}$
 equations (\ref{eom})  take the form:
\be\label{eom2}
\partial_{+}(\tilde{g}^{-1}\partial_{-}\tilde{g})=0\, ,\;\;\; {\rm Tr}(\tilde{g}^{-1}\partial_{-}\tilde{g}T_H)={\rm Tr}(\tilde{g}\partial_{+}\tilde{g}^{-1}T_H)=0\, .
\ee
%where $\tilde{g}=hgh^{-1}$.

The canonical symplectic form density, obtained  following the general prescription \cite{Crnkovic:1987tz,Crnkovic:1986ex,Gawedzki:1990jc},
is given by:
\be\label{densgh}
\Pi^{G/H}(g,h)=\Pi^G(\tilde{g})+\partial_x\Psi(h,g)\, ,
\ee

where
\be\label{formim}
\Psi(h,g)={\rm tr}\; h^{-1}d h(g^{-1}d g+d g g^{-1}+g^{-1}h^{-1}d h g)\, .
\ee

Some properties of the form (\ref{formim}) are summarized in appendix A.

Integrating   (\ref{densgh})  we get  the canonical symplectic form:
%\be
%\Omega^{G/H}=\int_0^{2\pi}\Pi^G(\tilde{g})+\Psi(h(2\pi),g(2\pi))-\Psi(h(0),g(0))
%\ee

%Using equation (\ref{om1}) in appendix  A we get
\be\label{ghfor}
\Omega^{G/H}={k\over 4\pi}\int_0^{2\pi}\Pi^G(\tilde{g})dx+{k\over 4\pi}\Psi(\rho^{-1}, hgh^{-1}(0))\, .
\ee
Collecting (\ref{dpi}), (\ref{mongrho}) and (\ref{om4}) one can show that the form (\ref{ghfor}) is closed.

Equations (\ref{eom2})  can be solved  in the terms of the chiral fields:
\be\label{newdec}
\tilde{g}=g_{L}(x^+)g_{R}^{-1}(x^-)\, ,\hspace{1cm} {\rm Tr}(\partial_yg_Lg_L^{-1}T_H)={\rm Tr}(\partial_yg_Rg_R^{-1}T_H)=0
\ee
with the monodromy properties:
\be\label{monpr}
g_L(y+2\pi)=\rho^{-1}g_L(y)\gamma\, ,  \hspace{1cm} g_R(y+2\pi)=\rho^{-1}g_R(y)\gamma\, .
\ee

The monodromy properties (\ref{monpr}) imply  that the chiral fields $g_{L,R}$ should be written as products of fields as well:

\be\label{decomppp}
g_{L}=h_B^{-1}g_A\, , \hspace{1cm}  g_{R}=h_D^{-1}g_C\, ,
\ee
where $h_B, h_D\in H$ and $g_A,g_C\in G$.
The fields in (\ref{decomppp}) should additionally satisfy:
\be\label{idenpr}
{\rm tr} [T_H(\partial_y h_B h_B^{-1}-\partial_y g_A g_A^{-1})]=0\, ,\hspace{1cm}{\rm tr} [T_H(\partial_y h_D h_D^{-1}-\partial_y g_C g_C^{-1})]=0
\ee
and
\be\label{monhgl}
h_B(y+2\pi)=h_B(y)\rho\, ,\hspace{1cm}g_A(y+2\pi)=g_A(y)\gamma\, ,
\ee
\be\label{monhgr}
h_D(y+2\pi)=h_D(y)\rho\, ,\hspace{1cm}g_C(y+2\pi)=g_C(y)\gamma\, .
\ee

Using (\ref{idenpr}) one can show:
\be\label{id33}
{\rm tr}[g_{L}^{-1}\delta g_{L}\partial_y (g_{L}^{-1}\delta g_{L})]={\rm tr}[ g_{A}^{-1}\delta g_{A}\partial_y (g_{A}^{-1}\delta g_{A})
- h_{B}^{-1}\delta h_{B}\partial_y (h_{B}^{-1}\delta h_{B})+\partial_y(\delta h_B h_B^{-1}\delta g_A g_A^{-1})]
\ee
and similarly for $g_R$ and $h_D$, $g_C$.

Collecting (\ref{newdec}), (\ref{monpr}), (\ref{decomppp}), (\ref{monhgl}), (\ref{monhgr}), (\ref{id33}) and  (\ref{denlr})
one can show that
\be\label{gaugsymp}
\Omega^{G/H}=\Omega^{\rm chiral}(g_A,\gamma)-\Omega^{\rm chiral}(g_C,\gamma)-\Omega^{\rm chiral}(h_B,\rho)+\Omega^{\rm chiral}(h_D,\rho)
\ee

Comparing (\ref{gaugsymp}) with (\ref{rldeco}) and remembering that the latter is the symplectic form 
of the Chern-Simons theory on ${\cal A}\times R$, we arrive at the conclusion that the phase space of
the gauged WZW model on a cylinder coincides with that of double Chern-Simons theory \cite{Gawedzki:2001ye,Moore:1989yh} on  ${\cal A}\times R$.

\section{Permutation branes}
Worldvolume $Q$ of the permutation branes on product of cosets  $G/H\times G/H$  corresponding to a  primary $(\mu,\nu)$   has been constructed in \cite{Sarkissian:2006xp}
and have the form:
\be\label{bibrane}
(g_1,g_2)=(cbp, Lp^{-1}L^{-1})\, ,
\ee
where 
$p\in G$, $L\in H$, $c\in C_H^{\nu}$, $b\in C_G^{\mu}$  and $C_G^{\mu}$  are  the conjugacy classes in $G$:
\be
C_G^{\mu}=\{\beta f_{\mu}\beta^{-1}=\beta e^{2i\pi\mu/k}\beta^{-1},\;\;\;\; \beta\in G\}\, ,
\ee
where $\mu\equiv${\boldmath $\mu\cdot H$} is a highest weight representation integrable at level $k$,
taking value in the Cartan subalgebra of the $G$ Lie algebra. $C_H^{\nu}$ are the similarly defined conjugacy  classes
in $H$. If $G$ and $H$ possess common center,  $\mu$ and $\nu$ should satisfy the selection rule \cite{Elitzur:2001qd}.

 To write the action  one  should 
introduce  an auxiliary disc $D$ satisfying the condition $\partial B=\Sigma+D$
and continue the fields $g_1$ and $g_2$ on this disc always holding 
the condition (\ref{bibrane}).

The action with the boundary condition (\ref{bibrane}) has the form
\be\label{actdefg}
S^{G/H\times G/H}_{\cal P}=S^{G/H}(g_1,A_1)+S^{G/H}(g_2,A_2)-{k\over 4\pi}\int_D\varpi(L,p,c,b)
\ee
where
\be
\varpi(L,p,c,b)=\Omega^{(2)}(c,b)-{\rm tr}((cb)^{-1}d(cb)dpp^{-1})+\Psi(L,p)\, ,
\ee
where
\be\label{ombig}
\Omega^{(2)}(c,b)=\omega_{\nu}(c)-{\rm tr}(c^{-1}dcdbb^{-1})+\omega_{\mu}(b)
\ee
and $\omega_{\mu}(C)$ is defined in (\ref{omfik}) and $\Psi(L,p)$ is defined in (\ref{formim}).
The form $\varpi(L,p,c,b)$ satisfies the condition:
\be\label{extdef}
d\varpi(L,p,c,b)=\omega^{WZ}(g_1)|_{Q}+\omega^{WZ}(g_2)|_{Q}\, .
\ee
One can check that the action (\ref{actdefg}) is invariant under the gauge transformations:
\bea\label{trang}
g_1&\rightarrow &h_1g_1h_1^{-1}\, ,\hspace{1cm} A_1\rightarrow h_1A_1h_1^{-1}-dh_1h_1^{-1}\, ,\\ \nonumber
g_2&\rightarrow &h_2g_2h_2^{-1}\, ,\hspace{1cm} A_2\rightarrow h_2A_2h_2^{-1}-dh_2h_2^{-1}\, ,
\eea
where $h_1: \Sigma\rightarrow H$, $h_2: \Sigma\rightarrow H$.
For this purpose note that under (\ref{trang}) the boundary parameters transform in the following way:
\be
 p\rightarrow  h_1ph_1^{-1}\, ,\;\;\; c\rightarrow  h_1ch_1^{-1}\, ,\;\;\;
b\rightarrow  h_1bh_1^{-1}\, ,\;\;\; L\rightarrow h_2Lh_1^{-1}\, .
\ee

The gauge invariance follows from the Polyakov-Wiegmann identities and the transformation properties 
of $\varpi(L,p,c,b)$:
\be\label{varptran}
\varpi(h_2Lh_1^{-1},h_1ph_1^{-1},h_1ch_1^{-1},h_1bh_1^{-1})-\varpi(L,p,c,b)\, ,
=-\Psi(h_1,cbp)-\Psi(h_2,Lp^{-1}L^{-1})
\ee
which can be obtained using formulae 
of appendix A.

Consider $G/H\times G/H$ product of coset models on the strip $R\times [0,\pi]$ with boundary conditions
on both sides given  by the permutation branes:

\be\label{bnc1}
(g_1,g_2)(0)=(C_2C_1p_1, L_1p_1^{-1}L_1^{-1})
\ee
\be\label{bnc2}
(g_1,g_2)(\pi)=(C_4C_3p_2, L_2p_2^{-1}L_2^{-1})
\ee
Here $C_1\in C_G^{\mu_1}$, $C_2\in C_H^{\mu_2}$, $C_3\in C_G^{\mu_3}$,$C_4\in C_H^{\mu_4}$,
$L_1,L_2\in H$, $p_1,p_2\in G$.

The boundary equation of motion resulting from the action (\ref{actdefg})  at $x=0$  are:

\be\label{dee1}
g_1^{-1}D_{-}g_1+L^{-1}_1g_2D_{+}g_2^{-1}L_1=0
\ee
\be\label{dee2}
C_2^{-1}g_1D_{+}g_1^{-1}C_2+L^{-1}_1g_2^{-1}D_{-}g_2L_1=0
\ee
\be\label{dee3}
L^{-1}_1D_tL_1=0\;\;\; C_2^{-1}D_tC_2=0
\ee
where $D_t=D_{+}-D_{-}$,   $D_{\pm}L=\partial_{\pm}L+A_{2\pm}L-LA_{1\pm}$, $D_{\pm}g_1=\partial_{\pm}g_1+[A_{1\pm},\, g_1]$,
$D_{\pm}g_2=\partial_{\pm}g_2+[A_{2\pm},\, g_2]$,  $D_{\pm}C_2=\partial_{\pm}C_2+[A_{1\pm},\, C_2]$.

Derivation of the equations (\ref{dee1}), (\ref{dee2}), (\ref{dee3}) is outlined in the appendix B.
Parameterising again flat gauge fields as 
\be
A_1=h_1^{-1}dh_1\hspace{1cm} A_2=h_2^{-1}dh_2
\ee
one can define as before
\bea
\tilde{g}_1=h_1g_1h_1^{-1}\hspace{1cm} \tilde{g}_2=h_2g_2h_2^{-1}\\ \nonumber
\tilde{C}_1=h_1C_1h_1^{-1}\hspace{1cm}\tilde{C}_2=h_1C_2h_1^{-1}\\ \nonumber
\tilde{p}_1=h_1p_1h_1^{-1}\hspace{1cm} \tilde{L}_1=h_2L_1h_1^{-1}\\ \nonumber
\tilde{C}_3=h_1C_3h_1^{-1}\hspace{1cm}\tilde{C}_4=h_1C_4h_1^{-1}\\ \nonumber
\tilde{p}_2=h_1p_2h_1^{-1}\hspace{1cm} \tilde{L}_2=h_2L_2h_1^{-1}
\eea
and we have the bulk equations (\ref{eom2}) for $\tilde{g}_1$ and $\tilde{g}_2$ and boundary equations take the form:
\be\label{g1per}
\tilde{g}_1^{-1}\partial_{-}\tilde{g}_1+\tilde{L}^{-1}_1\tilde{g}_2\partial_{+}\tilde{g}_2^{-1}\tilde{L}_1=0
\ee
\be\label{g2per}
\tilde{C}_2^{-1}\tilde{g}_1\partial_{+}\tilde{g}_1^{-1}\tilde{C}_2+\tilde{L}^{-1}_1\tilde{g}_2^{-1}\partial_{-}\tilde{g}_2\tilde{L}_1=0
\ee
\be\label{eomlper}
\tilde{L}_1^{-1}\partial_t\tilde{L}_1=0\;\;\; \tilde{C}_2^{-1}\partial_t\tilde{C}_2=0
\ee
Equation (\ref{eomlper}) implies that $\tilde{L}_1$ and $\tilde{C}_2$ are constant along the boundary.
Boundary conditions (\ref{bnc1}) and (\ref{bnc2}) imply
\be\label{bnc1til}
(\tilde{g}_1,\tilde{g}_2)(0)=(\tilde{C}_2\tilde{C}_1\tilde{p}_1,\tilde{L}_1\tilde{p}_1^{-1}\tilde{L}_1^{-1})
\ee
\be\label{bnc2til}
(\tilde{g}_1,\tilde{g}_2)(\pi)=(\tilde{C}_4\tilde{C}_3\tilde{p}_2, \tilde{L}_2\tilde{p}_2^{-1}\tilde{L}_2^{-1})
\ee
Using the chiral decomposition one can solve the boundary equation of motion
\be\label{solper1}
g_{1R}(y)=\tilde{L}_1^{-1}g_{2L}(-y)m
\ee
\be\label{solper2}
g_{2R}(y)=\tilde{L}_1\tilde{C}_2^{-1}g_{1L}(-y)n
\ee

Equations (\ref{solper1}) and (\ref{solper2}) indeed imply (\ref{bnc1til})
with
\be\label{p1}
\tilde{p}_1(t)=\tilde{C}_2^{-1}g_{1L}(t)ng_{2L}^{-1}(t)\tilde{L}_1
\ee
\be\label{c1}
\tilde{C}_1=\tilde{C}_2^{-1}g_{1L}(t)m^{-1}n^{-1}g_{1L}^{-1}(t)\tilde{C}_2
\ee
To have that $\tilde{C}_1\in C_G^{\mu_1}$ we should require $m^{-1}n^{-1}\equiv R_0\in  C_G^{\mu_1}$.

To satisfy (\ref{bnc2til}) we assume the following monodromy properties of $g_{1L}$ and $g_{2L}$
\be\label{monprdef}
g_{1L}(y+2\pi)=\rho_1^{-1}g_{1L}(y)\gamma_1\hspace{1cm}
g_{2L}(y+2\pi)=\rho_2^{-1}g_{2L}(y)\gamma_2
\ee
Now one can show that (\ref{bnc2til}) is satisfied with
\be\label{p2}
\tilde{p}_2(t)=\tilde{L}_2^{-1}\tilde{L}_1\tilde{C}_2^{-1}\rho_1g_{1L}(\pi+t)\gamma_1^{-1}ng_{2L}^{-1}(\pi+t)\tilde{L}_2
\ee
\be\label{c3}
\tilde{C}_3=\tilde{C}_4^{-1}g_{1L}(\pi+t)m^{-1}\gamma_2n^{-1}\gamma_1(\tilde{C}_4^{-1}g_{1L}(\pi+t))^{-1}
\ee
if we require
\be
\rho_2^{-1}=\tilde{L}_2\tilde{L}_1^{-1}
\ee
and
\be
\rho_1^{-1}\tilde{C}_2\tilde{L}_1^{-1}\tilde{L}_2=\rho_1^{-1}\tilde{C}_2\tilde{\rho}_2^{-1}=\tilde{C}_4
\ee
where $\tilde{\rho}_2=\tilde{L}_1^{-1}\rho_2\tilde{L}_1$.

To have that $\tilde{C}_3\in C_G^{\mu_3}$ we should require $m^{-1}\gamma_2n^{-1}\gamma_1=\tilde{\gamma}_2 R_{0}\gamma_1= R_{\pi}\in  C_G^{\mu_3}$, where
$\tilde{\gamma}_2=m^{-1}\gamma_2m$.

The monodromies (\ref{monprdef}) as before can be realized in the terms of the decomposition of the fields $g_{1L}$ and $g_{2L}$ 
as products:
\be
g_{1L}=h_B^{-1}g_A\, , \hspace{1cm}  g_{2L}=h_D^{-1}g_C
\ee
of the new fields $h_B$, $g_A$, $h_D$, $g_C$ possessing the monodromy properties:
\be
h_B(2\pi)=h_B(0)\rho_1\, ,\hspace{1cm}
g_A(2\pi)=g_A(0)\gamma_1\, ,
\ee
\be
h_D(2\pi)=h_D(0)\rho_2\, ,\hspace{1cm}
g_C(2\pi)=g_C(0)\gamma_2\, ,
\ee
and satisfying (\ref{idenpr}).

The symplectic form of product of the gauged WZW models on the strip with boundary conditions specified by the permutation branes can be written using the symplectic form density (\ref{densgh}) and the form $\varpi$:
\be\label{ghfrodef}
\Omega^{G/H}_{\cal P}={k\over 4\pi}\left[\int_0^{\pi}\Pi^{G/H}(g_1,h_1)dx+\int_0^{\pi}\Pi^{G/H}(g_2,h_2)dx+\varpi(g_1(0),g_2(0))-\varpi(g_1(\pi),g_2(\pi))\right]\, .
\ee
Substituting in (\ref{ghfrodef}) the symplectic form density (\ref{densgh})  and  using  the transformation property (\ref{varptran})
 we obtain:

\be\label{ghdef12}
\Omega^{G/H}_{\cal P}={k\over 4\pi}\left[\int_0^{\pi}\Pi(\tilde{g}_1)dx+\int_0^{\pi}\Pi(\tilde{g}_2)dx+\varpi( \tilde{L}_1, \tilde{p}_1,  \tilde{C}_2, \tilde{C}_1)-\varpi( \tilde{L}_2, \tilde{p}_2,  \tilde{C}_4, \tilde{C}_3)\right]\, ,
\ee
where $\tilde{p}_1$ and $\tilde{C}_1$ defined in (\ref{p1}) and (\ref{c1}) and $\tilde{p}_2$ and $\tilde{C}_3$ defined in (\ref{p2}) and (\ref{c3}).

Using (\ref{decomp})  one can obtain for (\ref{ghdef12}):
\be\label{om2perm}
\Omega^{G/H}_{\cal P}=\Omega^{LG}(s_1,\tau_1)+\Omega^{LG}(s_2,\tau_2)-\Omega^{LG}(s_3,\tau_3)-\Omega^{LG}(s_4,\tau_4)+\Omega_1^{\rm bndry}-\Omega_2^{\rm bndry}
\ee
\bea\label{bndryperm}
\Omega_1^{\rm bndry}={\rm tr}[(i\delta\tau_1)f_1^{-1}\delta f_1]
+{\rm tr}[(i\delta\tau_2)f_2^{-1}\delta f_2]\\ \nonumber
+{k\over 4\pi}\left[\omega_{\tau_1}(\gamma_1)+\omega_{\tau_2}(\tilde{\gamma}_2)+\omega_{\mu_1}(R_0)-
\omega_{\mu_{3}}(R_{\pi})\right.\\ \nonumber
-\left.{\rm tr}(R_0^{-1}\delta R_0\delta \gamma_1\gamma_1^{-1})
-{\rm tr}(\tilde{\gamma}_2^{-1}\delta\tilde{\gamma}_2\delta R_0R_0^{-1})-{\rm tr}(R_0^{-1}\tilde{\gamma}_2^{-1}\delta\tilde{\gamma}_2R_0\delta \gamma_1\gamma_1^{-1})\right]
\eea
\bea\label{bndryperm0}
\Omega_2^{\rm bndry}={\rm tr}[(i\delta\tau_3)f_3^{-1}\delta f_3]
+{\rm tr}[(i\delta\tau_4)f_4^{-1}\delta f_4]\\ \nonumber
+{k\over 4\pi}\left[\omega_{\tau_3}(\rho_1)+\omega_{\tau_4}(\tilde{\rho}_2)-\omega_{\mu_2}(\tilde{C}_2)+
\omega_{\mu_4}(\tilde{C}_4)\right.\\ \nonumber
+\left.{\rm tr}(\delta \tilde{C}_2\tilde{C}_2^{-1}\delta \rho_1\rho_1^{-1})
+{\rm tr}( \tilde{\rho}_2^{-1}\delta \tilde{\rho}_2\tilde{C}_2^{-1}\delta \tilde{C}_2)-{\rm tr}(\tilde{C}_2 \tilde{\rho}_2^{-1}\delta \tilde{\rho}_2\tilde{C}_2^{-1}\delta \rho_1\rho_1^{-1})\right]
\eea
Comparing (\ref{om2perm})  with the corresponding formulae in appendix C we arrive at the conclusion
that the phase space of product of coset  models on a strip with boundary conditions specified
by permutation branes is symplectomorphic to the phase space of the double Chern-Simons theory on an
annulus times the time-line and with  $G$ and $H$ gauge fields both coupled to two Wilson lines.

\section{Permutation branes in topological G/G coset}
In this  section we discuss permutation branes on the product of topological coset $G/G\times G/G$.

In the previous section we have seen that the phase space of the product of the gauged WZW models
on a strip with boundary conditions given by the permutation branes is symplectomorphic to the phase space of the
double Chern-Simons theory on an annulus  times the time-line with  $G$ and $H$ gauge fields both coupled to two Wilson lines. In the case when $G=H$ we arrive
at the conclusion that product of topological cosets $G/G\times G/G$ on a strip with boundary conditions
given by the permutation branes is equivalent to the Chern-Simons theory on the torus $T^2={\cal A}\cup (-{\cal A})$
times the time-line $R$ with four Wilson lines.
This can be verified by a direct calculation.
For the case $G=H$ the bulk equations of motion (\ref{eom2}) imply that $\tilde{g}_1$ and  $\tilde{g}_2$ are $(t,x)$ independent.

Therefore one has:

\be\label{gg1}
\tilde{g}_1(0)=\tilde{C}_2\tilde{C}_1\tilde{p}_1=\tilde{g}_1(\pi)=\tilde{C}_4\tilde{C}_3\tilde{p}_2
\ee
\be\label{gg2}
\tilde{g}_2(0)=\tilde{L}_1\tilde{p}_1^{-1}\tilde{L}_1^{-1}=\tilde{g}_2(\pi)=\tilde{L}_2\tilde{p}_2^{-1}\tilde{L}_2^{-1}
\ee

From equations (\ref{gg1}) and (\ref{gg2})  we get 

\be\label{mpc}
m\tilde{p}_2m^{-1}\tilde{p}_2^{-1}\tilde{C}_3^{-1}\tilde{C}_4^{-1}\tilde{C}_2\tilde{C}_1=1
\ee
where

\be
m=\tilde{L}_1^{-1}\tilde{L}_2
\ee

The symplectic form (\ref{ghdef12})  in this case reduces to 
\be\label{ghdef123}
\Omega^{G/G}_{\cal P}={k\over 4\pi}\left[\varpi( \tilde{L}_1, \tilde{p}_1,  \tilde{C}_2, \tilde{C}_1)-\varpi( \tilde{L}_2, \tilde{p}_2,  \tilde{C}_4, \tilde{C}_3)\right]\, ,
\ee

Comparing formulae (\ref{mpc}) and (\ref{ghdef123}) with the corresponding formulae in appendix C we arrive at the mentioned 
symplectomorphism of the product of topological cosets $G/G\times G/G$ on a strip with the boundary conditions given by the permutation branes
and that of Chern-Simons theory on the torus times the time-line  with four Wilson lines.
 
\section{Final Remarks}
The constructions in section 4 and 5 can be easily generalized to $N$-fold product of coset models $G/H$.
The ansatz for permutation branes has the form:
\be\label{peran}
(g_1,\ldots,g_N)=(C_2C_1p_{N-1}\cdots p_{1},L_1p_{1}^{-1}L_{1}^{-1},\ldots ,L_{N-1}p_{N-1}^{-1}L_{N-1}^{-1})
\ee 
where $C_1\in C_G^{\mu_1}$, $C_2\in C_H^{\mu_2}$, $p_i\in G$, $L_i\in H$.
The ansatz is invariant under the $N$-fold adjoint action : $g_i\rightarrow h_ig_ih_i^{-1}$, where $h_i: \Sigma\rightarrow H$.
Using the Polyakov-Wiegmann identity (\ref{pwwz}) it is straightforward to check the existence of the two-form
$\varpi_N$ satisfying the relation:
 \be
\sum_1^N \omega^{\rm WZ}(g_i)|_{\rm brane}=d\varpi_N
\ee
Performing the same steps as before we arrive at the conclusion, that 
the phase space
of the $N$-fold product of the gauged WZW model $G/H$  on a strip with boundary conditions given by permutation branes is symplectomorphic to the phase space of the double Chern-Simons theory on a sphere with $N$ holes
times the time-line and with  $G$ and $H$ gauge fields both coupled to two Wilson lines. For the special case of the toplogical coset $G/G$  we get, 
that  the phase space
of the $N$-fold product of the topological cosets $G/G$  on a strip with boundary conditions given by permutation branes is symplectomorphic to the phase space of  Chern-Simons theory on a Riemann surface of genus $N-1$
times the time-line with four Wilson lines.
\newpage

\appendix

\section{Useful formulae}
In this appendix we collect some useful properties of the two-form $\Psi(h,g)$ defined by formula (\ref{formim}).
%\be
%\Psi(h,g)={\rm tr}\; h^{-1}d h(g^{-1}d g+d g g^{-1}+g^{-1}h^{-1}d h g)
%\ee

\be\label{om1}
\Psi(hL,p)=\Psi(L,p)+\Psi(h,LpL^{-1})\, .
\ee
\be\label{om2}
\Psi(Lh^{-1}, hph^{-1})=\Psi(L,p)-\Psi(h,p)\, .
\ee
\be\label{om}
\omega_{\mu}(hCh^{-1})-\omega_{\mu}(C)=-\Psi(h,C)\, .
\ee
%where $C=kfk^{-1}$.

%where $C_1=k_1f_1k_1^{-1}$,  $C_2=k_2f_2k_2^{-1}$
\be\label{om3}
\Omega^{(2)}(hC_1h^{-1},hC_2h^{-1})-\Omega^{(2)}(C_1,C_2)=-\Psi(h,C_1C_2)\, ,
\ee

\be\label{om5}
\Psi(h,C_1C_2)=\Psi(h,C_1)+\Psi(h,C_2)+(\tilde{C}_1^{-1}d\tilde{C}_1d\tilde{C}_2\tilde{C}_2^{-1}-C_1^{-1}dC_1dC_2C_2^{-1})\, ,
\ee
where $\tilde{C}_1=hC_1h^{-1}$  and $\tilde{C}_2=hC_2h^{-1}$\, .

\be\label{om4}
\omega^{WZW}(hgh^{-1})-\omega^{WZW}(g)=-d\Psi(h,g)\, .
\ee

\section{Boundary Equations of motion}
Computing variation of the  action (\ref{actdefg})  one obtains for the boundary part:
\bea\label{dffff}
{\rm tr}[g_1^{-1}\delta g_1(g_1^{-1}\partial_{+}g_1+g_1^{-1}\partial_{-}g_1)]dt
+{\rm tr}[g_2^{-1}\delta g_2(g^{-1}\partial_{+}g_2+g_2^{-1}\partial_{-}g)]dt\\ \nonumber
+2{\rm tr}[\delta g_1g_1^{-1}A_{1-}-A_{1+}g_1^{-1}\delta g_1+(\delta g_2g_2^{-1}A_{2-}-A_{2+}g_2^{-1}\delta g_2)]dt+B=0\, .
\eea
The last term $B$ is one-form satisfying the relation:
\be\label{formb}
{\rm tr}(g_1^{-1}\delta g_1(g_1^{-1}d g_1)^2)+{\rm tr}(g_2^{-1}\delta g_2(g_2^{-1}d g_2)^2)-\delta\varpi=dB\, .
\ee
Recalling that the first two terms come from the equation
\be
\delta\omega^{WZ}=d[{\rm tr}(g^{-1}\delta g(g^{-1}d g)^2)]\, ,
\ee
we see that the existence of the one-form $B$ satisfying  (\ref{formb}) is a consequence of the equation  (\ref{extdef}).
Using (\ref{bibrane}) one can compute $B$ explicitly:
\bea\label{bexpo}
&&B=A_{\mu_1}(C_1)+A_{\mu_2}(C_2)+{\rm tr}[C_2^{-1}\delta C_2dC_1C_1^{-1}-\delta C_1C_1^{-1}C_2^{-1}dC_2-\\ \nonumber
&&\delta pp^{-1}(C_2C_1)^{-1}d(C_2C_1)
+(C_2C_1)^{-1}\delta(C_2C_1)dpp^{-1}-L^{-1}\delta Lp^{-1}dp+\\ \nonumber
&&p^{-1}\delta pL^{-1}dL-
L^{-1}\delta Ldpp^{-1}+\delta pp^{-1}L^{-1}dL
-L^{-1}\delta Lp^{-1}L^{-1}dLp+\\ \nonumber
&&L^{-1}\delta LpL^{-1}dLp^{-1}]\, .
\eea

The one-form $A_{\mu}(C)$ was defined in \cite{Elitzur:2000pq}:
\be
A_{\mu}(C)={\rm tr}[h^{-1}\delta h(f_{\mu}^{-1}h^{-1}dhf_{\mu}-f_{\mu}h^{-1}dhf_{\mu}^{-1})]\, ,
\ee

where $C=hf_{\mu}h^{-1}$, $f_{\mu}=e^{2\pi i\mu/k}$,

and satisfies:
\be
{\rm tr}(g^{-1}\delta g(g^{-1}d g)^2)|_{g=C}-\delta\omega_{\mu}(C)=dA_{\mu}(C)\, .
\ee

$A_{\mu}(C)$ satisfies also another important relation on the time-line:
\be\label{bndrmndr}
{\rm tr}[g^{-1}\delta g(g^{-1}\partial_{+}g+g^{-1}\partial_{-}g)]dt+A_{\mu}(C)={\rm tr}[2\delta hh^{-1}(\partial_{+}gg^{-1}-g^{-1}\partial_{-}g)]dt\, ,
\ee
where $g=C$.  Let us explain the meaning of this equation.

 The left hand side of the (\ref{bndrmndr}) is a particular case of  (\ref{dffff})
and describes boundary equation of motion of the WZW model with the boundary condition
specified by the conjugacy class $C$, while the right hand side proportional to $J_L+J_R$,
what is the condition for the diagonal chiral algebra preservation.

 Now, using  (\ref{bibrane}) and (\ref{bexpo}), one can show by a straightforwatd calculation, that (\ref{dffff})
implies  the equations (\ref{dee1}), (\ref{dee2}), (\ref{dee3}) in section 4.

\section{Symplectic forms of the moduli space of flat connections on a Riemann surface}

In this appendix we briefly review the symplectic phase space of the Chern-Simons theory on the three-dimensional manifold of
the form $M\times R$, where $M$ is two-dimensional Riemann surface, $R$ is time direction, with $n$ time-like
Wilson lines assigned with representations $\lambda_i$.
It was shown in \cite{Elitzur:1989nr,Witten:1988hf}  that the phase space of the Chern-Simons theory in such a situation 
is given by the moduli space of flat connections on the Riemann surface $M$  punctured at the points
$z_i$ where Wilson lines hit $M$,  with the holonomies around punctures belonging to the conjugacy classes
 $C_G^{\lambda_i}=\eta e^{2\pi i\lambda_i/k}\eta^{-1}$. Therefore denoting holonomies
around handles $a_j$ and $b_j$ by $A_j$ and $B_j$, and around punctures by $M_i\in  C_G^{\lambda_i}$ we arrive at the conclusion that the phase
space of the Chern-Simons theory is
\be
{\cal F}_{g,n}=G^{2g}\times \prod_{i=1}^n C_G^{\lambda_i}
\ee
subject to the relation
\be\label{relmod}
[B_{g}, A_{g}^{-1}]\cdots[B_{1}, A_{1}^{-1}]M_n\cdots M_1=I\, ,
\ee
where
\be
[B_j, A_j]=B_jA_jB_j^{-1}A_j^{-1}\, ,
\ee
and to the adjoint group action.

The symplectic form on ${\cal F}_{g,n}$ was derived in \cite{Alekseev:1993rj} and has the form:
\be\label{omms2}
\Omega_{{\cal M}_{g,n}}=\sum_{i=1}^n\Omega_{M_i} +\sum_{j=1}^{g}\Omega_{H_j}\, ,
\ee
where
\be
\Omega_{M_i}={k\over 4\pi} \omega_{\lambda_i}(M_i)
+{k\over 4\pi} {\rm tr}(K_{i-1}^{-1}\delta K_{i-1}K_{i}^{-1}\delta K_{i})\, ,
\ee
\bea
\Omega_{H_j}&=&{k\over 4\pi}\Psi(B_j,A_j)+{k\over 4\pi}({\rm tr}(K_{n+2j-2}^{-1}\delta K_{n+2j-2}K_{n+2j-1}^{-1}\delta K_{n+2j-1})\\ \nonumber
&+&
{\rm tr}(K_{n+2j-1}^{-1}\delta K_{n+2j-1}K_{n+2j}^{-1}\delta K_{n+2j}))\, ,
\eea
and where
\be
K_i=M_i\cdots M_1\hspace{1cm} i\leq n\, ,
\ee
\bea
K_{n+2j-1}=A_j[B_{j-1}, A_{j-1}^{-1}]\cdots[B_{1}, A_{1}^{-1}]K_n\, ,\\ \nonumber
K_{n+2j}=[B_{j}, A_{j}^{-1}]\cdots[B_{1}, A_{1}^{-1}]K_n\hspace{1cm} 1\leq j\leq g\, .
\eea
$K_0$ can be chosen to be equal to the unity element. According to (\ref{relmod})  also
\be
K_{n+2g}=I\, .
\ee
$\omega_{\lambda}(M)$ and $\Psi(B,A)$ are defined in equations (\ref{omfik}) and (\ref{formim}) correspondingly.

It was also proved in \cite{Alekseev:1993rj} that quantization of the moduli space ${\cal F}_{g,n}$
with the symplectic form (\ref{omms2}) leads to the space of $n$-point conformal blocks on a Riemann 
surface of the genus $g$.

The last piece of information which we need is the symplectic form on the moduli space of flat connections 
on the punctured sphere with holes $S^2_{n.m}$, where $n$ as before is number of punctures and $m$
is number of holes. It was argued in \cite{Elitzur:1989nr,Gawedzki:2001rm} that the corresponding  symplectic
form $\Omega_{S^2_{n,m}}$ is given by:
\be\label{decsn}
\Omega_{S^2_{n,m}}=\Omega_{S^2_{n+m,0}}+\sum_{i=1}^m\Omega_i^{\rm LG}\, ,
\ee
where $\Omega^{\rm LG}$ is defined in (\ref{omlg}) and its geometrical quantization 
leads to the integrable representation of the affine algebra $\hat{\rm g}$ at level $k$.

\end{document}